\documentclass[prl,showpacs, twocolumn]{revtex4}

\usepackage{graphicx}

\begin{document}

\title
{Layer-thickness effects in  quai-two dimensional electron liquids. 
 }

\author
{ M.W.C. Dharma-wardana
}
\email[Email address:\ ]{chandre.dharma-wardana@nrc.ca}
\affiliation{
Institute of Microstructural Sciences, National Research Council of Canada, Ottawa, Canada. K1A 0R6\\
}
\date{\today}
\begin{abstract}
We use a mapping of the  quasi-2D electron liquid to
 a classical fluid and use the hypernetted-chain equation inclusive of
bridge corrections, i.e., CHNC, to calculate the electron pair-distribution functions
and exchange-correlation energies as a function of the 
density, layer width, spin-polarization and
temperature. The theory is free of adjustable parameters
and is in good accord 
with recent effective-mass and spin-susceptibility
results for HIGFET 2-D layers.
\end{abstract}
%
\maketitle
%
The
 2-D electron systems (2DES) present in GaAs or Si/SiO$_2$
nanostructures
 access a
wide range of electron densities under controlled conditions,
providing a wealth of information\cite{krav} which is
of basic and technological importance.
The 2DES is in the x-y plane and also
has a transverse 
extension in the lowest sub-band of the
hetero-structure\cite{afs}. The higher subbands are 
 far above the Fermi energy and no upward excitations are
possible. 
Although the $z$-motion is confined to the lowest subband, realistic layers
may have widths of  $\sim600$ \AA$\,$ or more,
 and this corresponds to $\sim 6$ effective
atomic units in GaAs.
Recent experiments
and theory have focused on these
layer-thickness effects\cite{tutuc,tan,morsen,asgari,zhang}.
The physics of such quasi-2DES depends on the
density parameter $r_s$, the layer thickness
$w$ which labels the $z$-charge distribution, the spin-polarization
$\zeta$, and the temperature $T$.
The 2D density $n$ defines the
 mean-disk radius $r_s=(\pi n)^{-1/2}$ per electron,
expressed in effective atomic units which
depend on the bandstructure mass $m_b$ and the ``background''
dielectric constant $\epsilon_b$.
Although $r_s$ is the ``small parameter'' in perturbation theory (PT),
here it is simply the electron-disk radius
and PT is {\it not} used. 

Finite-width effects of the 2DES 
 arise also in the
quantum Hall effect\cite{ahm, dassarmaqh}, where
an  ``unperturbed-$g$'' approximation, which uses the
pair-distribution function (PDF) of the ideal 2DES
and the quasi-Coulomb potential $W(r)$ of the thick 2D layer
are used to calculate energies.

While diagrammatic methods propose  ``turn-key'' 
procedures for calculating many-body properties,
they work only for weakly coupled (small $r_s$) systems, where
RPA-like approximations may be used. Varied results can be 
obtained using various approximations which 
go ``beyond'' RPA. Unfortunately, an approximation
 successful with one
 property usually fails for  other properties.
Such methods lead to negative PDFs,
incorrect local-field corrections in the response functions,
and disagreement with the compressibility sum rule etc., and
incorrect predictions of spin-phase transitions (SPT) at quite
high densities. 
 In fact, alternative approaches were sought by
 Singwi,
Tosi et al. (STLS)\cite{stls}, and Ichimaru et al.\cite{ichimaru},
and also within the Feenberg-type correlated-wavefunction
methods\cite{campbell}. Most of the currently available
results for strongly-coupled
systems have been generated using correlated-wavefunction approaches
via Quantum Monte Carlo (QMC) simulations. However, QMC remains 
a strongly computer-intensive numerical method which is
best suited for the study of simple ``bench-mark'' systems.

 We have
recently introduced a
conceptually and numerically simple, adequately accurate
method for strongly-correlated quantum systems
at zero and finite temperatures, 
where the objective is to work with the PDF of the quantum
fluid, generated from a
 classical Coulomb fluid whose temperature $T_q$ is chosen to
reproduce the correlation energy of the original quantum fluid at $T=0$.
The classical PDFs are obtained from an integral equation
equivalent to a classical Kohn-Sham
equation where the correlation effects are captured as a sum of
hyper-netted-chain (HNC) diagrams and bridge diagrams. This method, known
as a {\it C}$\,$lassical mapping to an HNC form, i.e, CHNC,
 was applied to the 3D and
2D electron fluids\cite{prl1,prb,prl2,prl3},
 to dense hydrogen fluid\cite{hyd}, 
and also to the two-valley system in Si-MOSFETS\cite{2valley,eplmass}. 
In each case we showed that the
PDFs, energies, 
etc., obtained from CHNC were in excellent
agreement with comparable QMC results\cite{ssc,locf}.
Variants of the method have also been discussed by Bulutay and Tanatar, and
by Khanh and Totsuji\cite{buluty}.

The advantage of  CHNC is that it affords a simple, semi-analytic
theory for strongly correlated systems where QMC becomes prohibitive
or technically impossible to carry out. The classical-fluid model
allows for physically motivated treatments of complex issues
like three-body clustering etc., via the
statistical mechanics of hard-disk reference fluids\cite{rosenfeld}.
The disadvantage of the method,
typical of such many-body approaches, is that it is
currently an ``extrapolation'' method taking off from the results   of
a model fluid. For 2D systems, the fully spin-polarized
ideally-thin uniform fluid is the model fluid\cite{prl2}.
 As the CHNC method has been described in
previous work, we do not give a detailed account here.
This study is a simple, direct application of CHNC to
 the quasi-2D potential $W(r)=V(r)F(r)$, where $V(r)=1/r$
and $F(r)$ is a form-factor accounting for the modifications
 arising from the
thickness effect.

\begin{figure}
\includegraphics*[width=9.0cm, height=12.0cm]{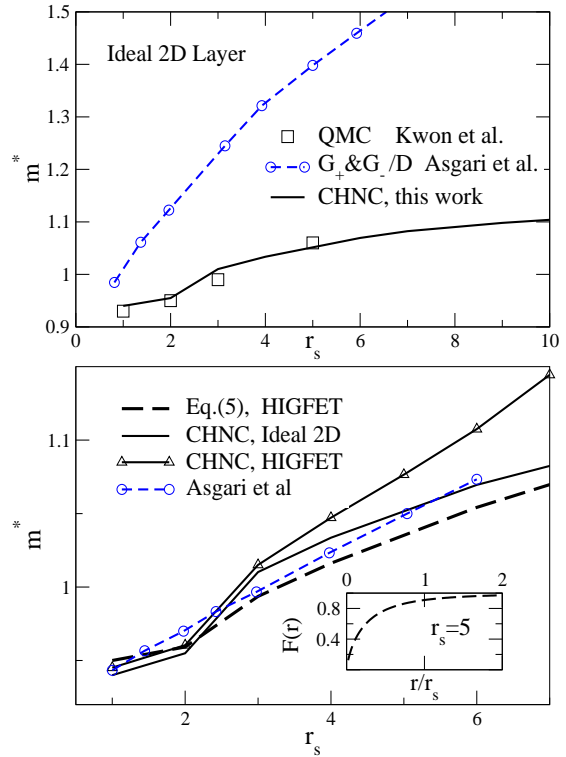}
\caption
{(a)The effective mass $m^*$ of an ideal 2D layer ($w=0$) obtained here are
 compared with the the QMC data of Ref.~\cite{kwon} and the
$G_+\&G_-/D$ calculation of Asgari et al\cite{asgari}. 
(b)The effective mass $m^*_H$ in a HIGFET using the CHNC ideal-2D
 finite-$T$ $g(r)$ in the ``unperturbed-$g$'' approximation, i.e.,
 Eq.~\ref{unpertg}. The HIGFET calculation of Asgari et al. is
 also shown.
}
\label{mstarfig}
\end{figure}

Most of our calculations are for the
 Fang-Howard(FH) approximation\cite{afs}
to the charge density $n(z)$ contained in HIGFET structures of the
type used by Zhu et al\cite{zhu}. In this case, $F(k)$, the
form factor in $k$-space,
 has an analytic form but $F(r)$ is numerically
 determined. It is technically convenient to work 
with an equivalent constant-density model (CDM)
 for which analytic forms are
available  for $F(r)$ as well as $F(k)$.
We present a potential $W(r,w)$ for a CDM which is
 {\it electrostatically equivalent} to the
the 2D potential  for any $n(z)$, e.g., the  FH-potential
defined by the parameter $b$.
The  method of replacing an inhomogeneous distribution by a
uniform distribution is suggested by the observation that the 
non-interacting total correlation function $h^0(r)=g^0(r)-1$ has
the form $\sim n(r)^2$,
where $n(r)$ is the
density-profile around the Fermi hole.
We replace the inhomogeneous  $n(z)$ by a slab of
constant-density  $n_{cd} $ which is easily shown to
accurately recover the  
 electrostatic potential of $n(z)$
in the 2-D plane. 
\begin{equation}
n_{cd}=1/w=\int n(z)^2dz
\end{equation}
Since the subband distribution is normalized to unity, the width
$w$ of the CDM is simply $1/n_{cd}$.
 Hence
quasi-2D layers can be labeled by their effective width $w$.
The CDM width $w$
for the $n(z)$ labeled by
the Fang-Howard $b$  is $w=16/(3b)$, and differs from 
the commonly quoted ``thickness'', $3/b$.
This provides an explicit example of the replacement of an inhomogeneous
distribution by a homogeneous distribution, already suggested 
in Ref.~\cite{ggsavin} and used for 2-electron atoms.
The quasi-2D potential for a CDM
 of width $w$ is given by
\begin{eqnarray}
W(r)&=&V(r)F(s),\;\; s=r/w, \;\; t=\surd{(1+s^2)} \nonumber \\
F(r)&=&2s[\log(t/s)+t]
\end{eqnarray}
This tends to $1/r$ for large $r$. 
The short-range behaviour is logarithmic,
and weaker than the $1/r$ potential.
The form factor $F(r)$ for a HIGFET with $r_s=5$ is shown in
the inset to Fig.~\ref{mstarfig}.
The
$k$-space form of the CDM potential is:
\begin{eqnarray}
W(k,w)&=&V(k)F(p),\,\,\, p=kw\\ 
F(p)&=&(2/p)\{(e^{-p}-1)/p+1\}
\end{eqnarray}
The form factors $F(s)$ and $F(p)$ tend to unity as $w\to 0$.
These $r$-space and $k$-space analytic forms of the CDM  
indirectly lead to analytic formulae for the FH from.
The $W(r)$ of the CDM is equivalent to that from the
original inhomogeneous distribution, and only
 $W(r)$  enters into the exchange-correlation and $g(r)$
calculations.
 In the case of GaAs-HIGFETS, if the depletion density could be
neglected\cite{morsen}, $r_s$
specifies the $b$ parameter and hence the width $w$ of the CDM. 
Then $b^3=33/(2r_s^2)$  and $w=2.09494r_s^{2/3}$. When the 
 exchange-correlation energy $E_{xc}(b,r_s)$ is included,
$b$ changes to $b^*$. The correction is $\sim 2-3$\% at low $r_s\sim 1$ and
decreases as $r_s$ increases. We have used $b^*$ in our
fully self-consistent CHNC calculations.

{\it Exchange and correlation in quasi-2D layers.}---
 The 
  noninteracting-2D  correlation function, $h^0(r)=g^0(r)-1$, 
  yields exact exchange energies for arbitrary 
$w$ as well as for $w=0$ (the ideal 2DES). On the other hand, the 
 correlation-energy evaluation needs
 the PDFs of the
 quasi-2D potential $\lambda W(r,w)$, at many values 
of the coupling constant $\lambda$. These  $g(r,\zeta,w,\lambda)$
can be calculated using the CHNC. 
However, the use of the {\it unperturbed}-$g$ 
approximation
used in Quantum Hall studies\cite{ahm}
can be useful here too.
 De Palo
  et al.\cite{morsen} have in fact exploited  such an
 approach where the $g(r,\zeta,w=0)$
 of the ideally thin layer are used to calculate a correction energy
  $\Delta$ given by,
  \begin{equation}
  \label{unpertg}
  \Delta  =(n/2)\int 2\pi rdr [W(r)-V(r)]h(r,\zeta,w=0)
  \end{equation}
  Then the total  $E_{xc}(r_s,\zeta,w)$ is
  obtained by adding to $\Delta$ 
  the known $E_{xc}$ of the ideally-thin  system. 
  The ``unperturbed'' $g(r)$ needed in Eq.~\ref{unpertg}
   at $T=0$ are the QMC $g(r)$\cite{ggpair}.
Equation~\ref{unpertg}
  can also be applied to  the finite-$T$ ideal 
  $g(r_s,\zeta,T)$ obtainable from  CHNC. However,
 this approach is found to be insufficient  for calculating $m^*$.   
 
 De Palo et al\cite{morsen} have performed Diffusion Monte Carlo 
simulations at $r_s$ =5 for HIGFETS with $b=0.8707$, i.e,
 a CDM width  $w$=6.1256 a.u., and find that
the error compared to the full simulation is about
2\%. Since the ratio $w/r_s$ decreases
 in the  HIGFET  as $r_s$ increases, 
the HIGFET approaches the thin-layer model for large $r_s$.
Hence Eq.~\ref{unpertg} is satisfactory for $r_s\ge $ 5, and 
unreliable for small $r_s$, e.g., below $r_s=3$.
Equation~\ref{unpertg} neglects the renormalization of the
kinetic energy, correction  of $b$ to $b^*$,
 as well as the changes in $g(r)$ due to the
changed potential. 
We have used both the full CHNC which accounts for
all these effects, and also
the ``unperturbed-$g$'' approximation, Eq.~\ref{unpertg},
 and find that the latter is indeed satisfactory
for the calculation of $E_{xc}$  and the susceptibility enhancement
$m^*g^*=\chi_s/\chi_P$,
where  $\chi_s$, $\chi_P$  are the interacting and Pauli spin susceptibilities.
\begin{figure}
\includegraphics*[width=9.0cm, height=12.0cm]{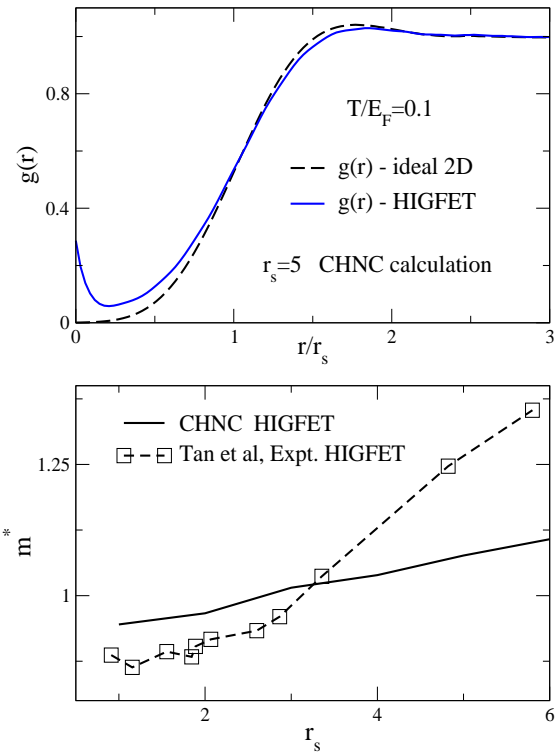}
\caption
{(a)The spin-averaged $g(r)$ at $r_s=5$
and $T/E_F$=0.1,
for an ideally thin layer, and for a HIGFET  calculated using CHNC.
(b)The effective mass $m^*_H$ in a HIGFET using the fully self-consistent
CHNC calculations for the energy at finite-$T$. 
 The experimental data are from
 Tan et al\cite{tan}.
}
\label{scfmstr}
\end{figure}

{\it Correlation energy at finite temperatures.}---
 The correlation contribution to the Helmholtz free energy of
 ideal ($w=0$) or thick layers ($w>0$) is readily calculated
 via Eq.~\ref{unpertg}, using the 
 CHNC-finite-$T$ $g(r)$. 
For example, consider $r_s=5$, and $T/E_F$=0.05, with 
  $\eta=0.3718$, i.e., the packing fraction of  the hard-disk fluid used
 to mimic the three-body  terms in the extended HNC equation. Then
$E_{xc}$ for the ideal 2D at $\zeta=0$ is -0.16902 a.u., while the HIGFET
$E_{xc}$  from Eq.~\ref{unpertg} and from the full
CHNC are -0.11197 a.u. and  -0.11467 a.u. respectively.
 As  discussed in earlier work\cite{ssc,eplmass},
  the exchange free energy $F_x$ and the correlation 
free energy $F_c$ at very
  low $T$ contain logarithmic terms which cancel with
  each other, so that the sum $F_{xc}=F_x+F_c$ is free of such
  terms. 
  At $r_s=5$ the cancellation is good to about 75\%, and
  this improves as $r_s$ increases. Although the two-component fluid (up
  and down spins) involves three distribution functions, we have,
  as before\cite{prl2}, used only one hard-disk bridge function, $B_{12}$, as
  clustering effects in $g_{ii}$ are mostly suppressed by the Pauli-exclusion. 
 This is a satisfactory approximation for $r_s \ge 3$, and for smaller $r_s$ 
 $B_{12}$ becomes comparable to $B_{ii}$, but negligible.
 Also,  $B_{12}$ is assumed to be independent of $\zeta$ and identical to the
 ideal- 2D ($w=0$) bridge function.
 
%
 %
%
%

{\it The spin-susceptibility, $m^*$  and $g^*$.}---
The results for the  $F_{xc}(rs,\zeta,T,w)$ of the
 ideal or thick 2DES suffice
to calculate the spin-susceptibility
 enhancement,
the effective mass $m^*$ and the effective Land\'{e} factor $g^*$.
 We calculate the
quantity $A=1+B(x)$, where $B(x)$ is the ratio
 of the second derivative with respect to $x$
of $F_{xc}(x)$, and $F_0(x)$, where $x$ is $\zeta$ for the
spin susceptibility calculation, and $x=T$ for the $m^*$ calculation.
$F_0(x)$ is the noninteracting Helmholtz free energy. 
  The CHNC is used to obtain any 
data
 (e.g, at finite -$T$)  unavailable from QMC.
 The effective mass $m_H^*$ obtained for the HIGFET using 
Eq.~\ref{unpertg}, where the ideal 2D-finite-$T$ $g(r)$ is used,
is presented
 in the lower panel of 
Fig.~\ref{mstarfig}.
  The upper panel shows the ideal 2D-layer $m^*$, 
  in good agreement with the four QMC values.
 This  contrasts the $m^*$ proposed
 by Asgari et al,  denoted 
 $G_+\&G_-/D$ in their paper\cite{asgari}.
 A study of the local-field factor of the 
 2D response function\cite{locf}
 shows that the formation of singlet-pair correlations is
 complete by $r_s\sim 5$, and after that the structure of the fluid
 remains more or less unchanged, until the spin-phase transition (SPT)
 is reached. The
 rapid  rise in $m^*$
 before $r_s\sim 5$ and the subsequent slow-down are probably related to 
 the formation and persistence of  singlet structure in the 2D fluid.
 
 In the lower panel of Fig.~\ref{mstarfig} we present the 
 effective mass $m^*_H$ of the electrons in the HIGFET. 
The ``unperturbed-$g$'' model, Eq.~\ref{unpertg},
 is unable to approach the
experimental results (Fig.~\ref{scfmstr}) of Tan et al. 
Hence we repeat the $m^*$ calculations
using the fully self-consistent $g(r)$ from CHNC. The
 $g(r)$ for the ideal 2DES  and the fully self-consistent 
$g(r)$ for a HIGFET at $r_s=5$, $T/E_f=0.1$ are shown
in the upper panel of Fig.~\ref{scfmstr}. 
\begin{figure}
\includegraphics*[width=8.0cm, height=10.0cm]{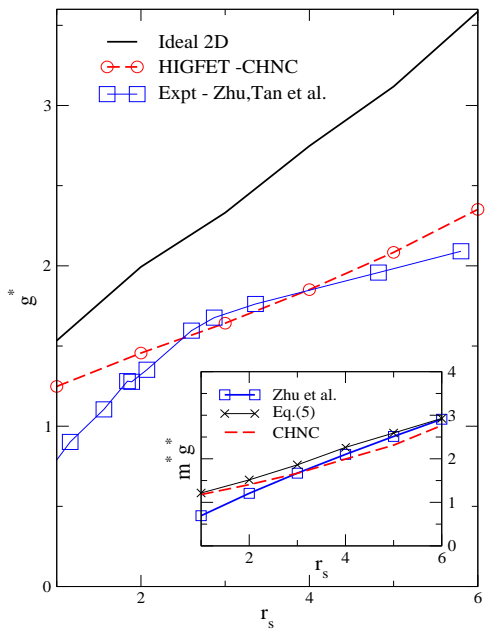}
\caption
 { The Land\'e $g$-factor for the ideal 2D Layer is obtained
from the QMC $m^*g^*$ of ref.~\cite{attac}, divided by the CHNC ideal 2D $m^*$.
The experimental HIGFET $g_H^*$ (boxes) is from the $m^*g^*$ of Zhu et al.,
divided by the $m^*$ of Tan et al. We also show $g_H^*$ calculated from
the $m^*g^*$ and the $m_H^*$ results for the HIGFET.
The inset shows the spin-susceptibility enhancement $m^*g^*$ from the
$\zeta$ dependent energies calculated from Eq.~\ref{unpertg}, where
the ideal 2D $g(r)$ is used, and 
from the full CHNC calculation using the $g(r)$ consistent with
the quasi-2D potential. 
}
\label{gstarfig}
\end{figure}
The weakened repulsion
in $W(r)$ for small-$r$
in the
quasi-2DES allows easier electron clustering, 
strongly boosting the  $g(r)$ near $r=0$. The HIGFET $m^*$, 
 evaluated from the
second derivative of  $F_{xc}(T)$ in the 
range $T/E_F=0.05$ to 0.1,  is shown in the lower panel.
Note that $m^*$ in our work {\it increases}
 from the ideal ($w=0$) system to the
 HIGFET, where as Asgari et al. predict the opposite.
{\it Enhancement of the spin susceptibility.}
De Palo et al\cite{morsen}. have calculated $m^*g^*$ from the 
ideal QMC $g(r)$, using the ``unperturbed-$g$'' approximation,
 and shown that they obtain
close agreement with the data for very narrow 2-D systems\cite{vakili}
as well as for the thicker 2DES in HIGFETS\cite{zhu}, as shown
in the inset to  Fig.~\ref{gstarfig}.
We obtain results in close agreement when the ``unperturbed-$g$''
approximation, Eq.~5, is used.
 The $m^*g^*$ from the fully self-consistent $g(r)$ are
also shown in the inset, and shows a deviation of $\sim 5$\%. 
This deviation is probably a short-coming of the CHNC model,
associated with the use of a spin-independent bridge function.
%

{\it The effective Lande\'e-$g$ factor.}---
	The agreement between the experimental $\chi_s/\chi_P$ ratio (i.e,
$m^*g^*$) and the theoretical results, especially those from
QMC for the ideal 2DES and the HIGFET suggests
that $m^*g^*$ is known with some confidence. Hence we
may  extract the
effective Land\'e-$g$ factors for the ideal 2DES and the
 HIGFET (see Fig.~\ref{gstarfig}), using the available $m^*$ values.
The strong increase in $m^*$ with $r_s$ in HIGFETS implies that the
$g^*$-factor is less sensitive to $r_s$.
  This
is similar to the behaviour of the two-valley system found in Si-MOSFETS
where there is {\it no} SPT\cite{2valley,eplmass}.
The additional inter-valley Coulomb interactions in MOSFETS weakens the role of
 exchange and enhanced the effect of singlet-pair  cluster
effects, leading to a strong increase around $r_s=5$. In the HIGFET, the
the weakening of the repulsive interaction increases
clustering, and boosts $m^*$, since quasi-particles
no longer move freely, but have
 to  drag a cloud of electrons associated with the enhanced
  $g(r)$ near $r/r_s\to 0$.
  This shows that the $m^*$ calculation is very sensitive to the accurate 
  evaluation of the ``on-top'' value of $g(r=0)$. In the CHNC, clustering and 
  such effects are controlled by the hard-disk bridge function, and the
  diffraction correction (de Broglie momentum $k_{th}$\cite{prl2})
  used to describe the quantum-scattering of two
  electrons. In this work we have simply used the $\eta$ and $k_{th}$ of the
  ideal 2DES, i.e., $w=0$. 
 Better agreement of $m^*$ with experiment would require
  an evaluation of these as a function of $w$. Similarly, complete QMC
 runs for finite $w$ would require back-flow and
three-body functions different from the usual RPA-like treatment of the 2DES. 

{\it Conclusion}-- We have used a single theoretical framework,
 i.e., the CHNC, with no 
parameters other than those previously used
 for the ideal 2D system\cite{prl2}, to
calculate the $F_{xc}(r_s,\zeta,T,w)$, and hence the $m^*$ and $g^*$
 of thick 2D layers. The results suggest that the enhanced mass in HIGFETS
 arises from the strong short-ranged correlations created by
 the weakened Coulomb repulsion due to the thickness effect.


\begin{thebibliography}{99}
%
\bibitem{krav}
S. V. Kravchenko and M. P. Sarachik, Rep. Prog.Phys. {\bf 67}, 1 (2004)
\bibitem{afs}
T. Ando, B. Fowler, and F. Stern, Rev. Mod. Phys. {\bf 54}, 437 (1982)
\bibitem{martin}
Y-H Kim, I-H Lee, S. Nagaraja, J-P Leburton, R. Q. Hood, R. M. Martin,
 Phys. Rev. B {\bf 61}, 5202 (2000)
\bibitem{ahm}
A. H. MacDonald and G. C. Aers, Phys. Rev. B {\bf 29}, 5976 (1984)
\bibitem{dassarmaqh}
F. C. Zhang and S. Das Sarma, Phys. Rev. B {\bf 33}, 2903 (1986)
\bibitem{tutuc}
E. Tutuc, S. Melinte, E. P. de Poortere, M. Sheyegan, and R. Winkler.
cond-mat/0301027
\bibitem{tan}
Y.-W Tan, J. Zhu, H. L. Stormer, L. N. Pfeiffer, K. N. Baldwin, and
K. W. West, Phys. Rev. Lett. {\bf 94},16405, (2005)
\bibitem{zhu}
J. Zhu, H. L. Stormer, L. N. Pfeiffer, K. W. Baldwin, and K. W. West,
Phys. Rev. Lett. {\bf 90}, 56805 (2003)
\bibitem{morsen}
S. De Palo, M. Botti, S. Moroni, and G. Senatore, cond-mat/0410145
\bibitem{asgari}
R. Asgari, B. Davoudi, M. Polini, M. P. Tosi, G. F. Giuliani, and
G. Vignale, cond-mat/0412665
\bibitem{zhang}
Y. Zhang and S. Das Sarma, cond-mat/0312565
\bibitem{stls}
 K. S. Singwi, M. P. Tosi,  R. H. Land, and A. Sjolander,
 Phys. Rev B {\bf 176}, 589 (1968) 
\bibitem{ichimaru}
S. Ichimaru, Rev. Mod. Phys. {\bf 54}, 1017 (1982)
\bibitem{campbell}
C. E. Campbell and J. G. Zabolitsky, Phys. Rev. B {\bf 27}, 7772 (1983), and
references therein.
\bibitem{prl1}
M. W. C. Dharma-wardana and F. Perrot, Phys. Rev. Lett. {\bf 84}, 959 (2000)
\bibitem{prb}
Fran\c{c}ois Perrot and M. W. C. Dharma-wardana, Phys. Rev. B, {\bf 62}, 14766 (2000)
\bibitem{prl2}
Fran\c{c}ois Perrot and M. W. C. Dharma-wardana,  Phys. Rev. Lett. {\bf 87},
 206404 (2001)
\bibitem{prl3}
M. W. C. Dharma-wardana and F. Perrot., Phys. Rev. Lett. {\bf 90}, 136601 (2003)
\bibitem{hyd}
M. W. C. Dharma-wardana and F. Perrot, Phys. Rev. B, {\bf 66}, 14110 (2002)
\bibitem{2valley}
M. W. C. Dharma-wardana and F. Perrot, Phys. Rev. B {\bf 70}, 035308 (2004)
\bibitem{ssc}
M. W. C. Dharma-wardana, Solid State Com. {\bf 127}, 783-788 (2003); the preliminary results for $m^*$ etc. reported here are superceded by the present report.
\bibitem{eplmass}
M. W. C. Dharma-wardana, Europhys. Lett. {\bf 67}, 552 (2004)
\bibitem{shash}
A. A. Shashkin, M. Rashmi, S. Anissimova, and S. V. Kravchenko, V. T. Dolgopolov,
T. M. Klapwijk, Phys. Rev. Lett. {\bf 91}, 46403 (2003)
\bibitem{locf}
M. W. C. Dharma-wardana and F. Perrot, Europhys. Lett.
{\bf 63}, 660-666 (2003)
\bibitem{buluty}
C. Bulutay and B. Tanatar, Phys. Rev. B {\bf 65}, 195116 (2002)
N. Q. Khanh and H. Totsuji, Solid State Com., {\bf 129},37 (2004)
\bibitem{rosenfeld}
Y. Rosenfeld, Phys.Rev. A {\bf 42}, 5978 (1990)
\bibitem{ggsavin}
P. Gori-Giorgi and A. Savin, cond-mat/0411179.
\bibitem{pdw84}
F. Perrot and M. W. C. Dharma-wardana, Phys. Rev. A {\bf 30}, 2619 (1984)
\bibitem{depalo}
We thank Stephania de Palo and Gaetano Senatore for
 for providing us
with their correction energy $\Delta$ and the full
 diffusion-QMC correction at
$r_s=5$. 
\bibitem{ggpair}
P. Gori-Giorgi, S. Moroni and G. B. Bachelet, Phys. Rev. B {\bf 70}, 115102 (2004)
\bibitem{kwon}
Y. Kwon, D. M. Ceperley, and R. M. Martin, Phys. Rev. B {\bf 50}, 1684 (1994)
\bibitem{vakili}
K. Vakili, Y. P. Shkolnikov, E. Tutuc, E. P. de Poortere, and M. Shayegan,
Phys. Rev. Lett. {\bf 92},226401 (2004)
\bibitem{attac}
C. Attaccalite, S. Moroni, P. Gori-Giorgi, and G. B. Bachelet, Phys. Rev.
Lett. {\bf 88}, 256601 (2002)
\end{thebibliography}
\end{document}